\documentclass{pnastwo}

\usepackage[dvips]{graphicx}
\usepackage{amssymb,amsfonts,amsmath,cite}

\begin{document}

\conflictofinterest{The authors declare no conflict of interest}


\footcomment{Abbreviations: FBA, flux balance analysis; MOMA,
minimization of metabolic adjustment}


\title{Identifying essential genes in {\it E. coli} from a metabolic optimization principle}

\author{Carlotta Martelli\affil{1}{Dipartimento di Fisica,
Universit\`a di Roma ``La Sapienza'', p.le A. Moro 2, 00185 Roma
(Italy)}, Andrea De Martino\affil{2}{CNR(ISC)/INFM(SMC) and
Dipartimento di Fisica, Universit\`a di Roma ``La Sapienza'', p.le
A. Moro 2, 00185 Roma (Italy)}, Enzo Marinari\affil{3} {Dipartimento
di Fisica, CNR/INFM and INFN, Universit\`a di Roma ``La Sapienza'',
p.le A. Moro 2, 00185 Roma (Italy)} , Matteo Marsili\affil{4}{The
Abdus Salam ICTP, Strada Costiera 11, 34014 Trieste (Italy)}, \and
Isaac P\'{e}rez Castillo\affil{5}{Department of Mathematics, King's
College London, Strand, London WC2R 2LS (UK)}}

\contributor{Submitted to Proceedings of the National Academy of
Sciences of the United States of America}

\maketitle

\begin{article}

\begin{abstract} Understanding the organization of reaction fluxes in
cellular metabolism from the stoichiometry and the topology of the
underlying biochemical network is a central issue in systems biology.
In this task, it is important to devise reasonable approximation
schemes that rely on the stoichiometric data only, because full-scale
kinetic approaches are computationally affordable only for small
networks (e.g. red blood cells, about 50 reactions). Methods commonly
employed are based on finding the stationary flux configurations that
satisfy mass-balance conditions for metabolites, often coupling them
to local optimization rules (e.g. maximization of biomass production)
to reduce the size of the solution space to a single point. Such
methods have been widely applied and have proven able to reproduce
experimental findings for relatively simple organisms in specific
conditions. Here we define and study a constraint-based model of
cellular metabolism where neither mass balance nor flux stationarity
are postulated, and where the relevant flux configurations optimize
the global growth of the system.  In the case of {\it E. coli}, steady
flux states are recovered as solutions, though mass-balance conditions
are violated for some metabolites, implying a non-zero net production
of the latter. Such solutions furthermore turn out to provide the
correct statistics of fluxes for the bacterium {\it E. coli} in
different environments and compare well with the available
experimental evidence on individual fluxes. Conserved metabolic pools
play a key role in determining growth rate and flux variability.
Finally, we are able to connect phenomenological gene essentiality
with `frozen' fluxes (i.e. fluxes with smaller allowed variability) in
{\it E. coli} metabolism.
\end{abstract}

\keywords{metabolic networks | optimization | conserved moieties |
gene essentiality}

\dropcap{C}ellular metabolism involves a complex network of
interactions and cross-regulations between metabolites, proteins and
genes. While our knowledge of regulatory functions at the genetic
level and at the level of protein-protein interactions is still in its
infancy, the biochemical network of reactions that describes
metabolism has been characterized in detail for many organisms
\cite{Pals1, Pals2, Pals3}. Most information about the metabolic
network is contained in the matrices $\mathbf{B}=\{b_i^\mu\}$ and
$\mathbf{A}=\{a_i^\mu\}$ describing, respectively, the input and
output stoichiometric coefficients of each metabolite $\mu$ (ranging
from $1$ to $P$) for all the metabolic reactions $i$ ($1$ to $N$).
Their knowledge allows for the definition of constraint-based models
from which a prediction of metabolic fluxes is possible
\cite{Pals4,Kau,Segre}.  These models typically rely on a steady-state
assumption that reflects the timescale separation between the (faster)
equilibration of metabolic processes and the (slower) dynamics of
genetic regulation. Under this condition, the concentration $ X^\mu$
of metabolite $\mu$ and the flux $\nu_i\geq 0$ of reaction $i$ are
constant in time and globally linked by a set of mass-balance
conditions:
\begin{equation}\label{fba}
\frac{{\rm d}X^\mu}{{\rm d}t}=\sum_{i=1}^N (a_i^\mu-b_i^\mu) \nu_i\equiv 0
 \quad,\quad\forall \mu=1,\ldots,P\;
\end{equation}
or, in matrix notation,
$(\mathbf{A}-\mathbf{B})\boldsymbol{\nu}=\mathbf{0}$, where
$\boldsymbol{\nu}=\{\nu_1,\ldots,\nu_N\}$ is a flux vector. The problem is
that of finding a flux vector satisfying the set of $P$ linear
equations (\ref{fba}). For real metabolic networks, the above system
is usually under-determined as $N>P$ (e.g. {\it E. Coli} has around
$1100$ reactions and less than $800$ metabolites \cite{Pals1}), so
that multiple solutions exist and one has to specify which of these
are the working, physical states of the network.

In organisms with high functional specificity or whose main objective
is to produce certain specified metabolites, like e.g.  red blood
cells, physical states are taken to be all those
consistent with the mass-balance conditions. In such cases, it is
important to be able to explore the solution space of (\ref{fba})
uniformly. Uniform sampling can be achieved in small networks with
Monte Carlo methods \cite{wiback} and, more recently, message-passing
algorithms have been employed \cite{bz,pag}. However computational
considerations still prevent the application of such techniques to
explore more general aspects of larger metabolic networks.

For more complex organisms, one normally complements (\ref{fba}) with
the assumption that the physical state of the network obeys a specific
optimization principle. In flux-balance analysis (FBA, \cite{Kau})
the choice usually falls on the maximization of biomass production, a
useful proxy of the growth capabilities of an organism. FBA has been
widely applied to different organisms (mostly bacteria) to investigate
general aspects of metabolism, like flux distributions in different
environments \cite{Bara1}, evolutionary dispensability of enzymes
\cite{Pal}, or the plasticity of the reaction network
\cite{Almaas}. Minimization of metabolic adjustment (MOMA,
\cite{Segre}) is instead able to predict the re-organization of fluxes
after a reaction knockout by minimizing the distance between the FBA
solution and that obtained from (\ref{fba}) after the knockout.

Here we consider a constraint-based model of metabolic activity with
the aim of characterizing flux states corresponding to optimal net
metabolic production. We neither assume constancy-in-time of fluxes,
nor do we impose mass-balance constraints on metabolites.  Rather, we
allow for these to be recovered as properties of the solutions.  This
assumption turns out to be able to reproduce the empirical statistics
of fluxes for the bacterium {\it E. coli} in different environments
with a remarkable accuracy, and the physical and biochemical origin of
the robustness of the emerging picture can be completely
understood. Further biological insight can be gained by comparing the
variability of individual fluxes with data on phenomenological gene
essentiality.

\section{Model definitions}

The abstract setup we consider was originally introduced by J. Von
Neumann to model growth in production economies as an autocatalytic
process \cite{VN}. We consider a system of $N$ chemical reactions
between $P$ reagents, with fluxes and concentrations evolving in
discrete time $t=0,1,\ldots$ (we leave the timescale unspecified).
For now, the system is assumed to be purely autocatalytic, so that the
total input of any given metabolite at a certain time step must come
from (all of or part of) the output at the previous time step. Let
$S_i(t)$ denote the scale at which reaction $i$ operates at time $t$,
so that the total input and output of metabolite $\mu$ are given by
$I^\mu(t)=\sum_i S_i(t) b_i^\mu$ and $O^\mu(t)=\sum_i S_i(t) a_i^\mu$,
respectively. Stability requires that
$C^\mu(t):=O^\mu(t)-I^\mu(t+1)\geq 0$ at all times, otherwise the
system would need an outside metabolic source to survive. We focus our
attention on the feasibility of dynamical paths with constant growth
rate $\rho>0$, where $I^\mu(t+1)=\rho\; I^\mu(t)$.  It is easy to see
that, in such paths, reaction rates must behave as $S_i(t)=s_i\rho^t$,
with constants $s_i\geq 0$ satisfying the $P$ linear constraints
\begin{equation}\label{vn}
c^\mu\equiv\sum_{i=1}^N (a_i^\mu-\rho b_i^\mu)s_i\geq 0
\quad,\quad \forall \mu=1,\ldots,P\; .
\end{equation}
The main assumption now is that the network's physical state ${\bf
s^\star}=\{s_i^\star\}$ corresponds to one of those for which the
growth rate $\rho$ attains its maximum possible value $\rho^\star$
compatible with the constraints (\ref{vn}).  The rationale for this
choice lies in the fact that, under general conditions, it can be
proven that paths with the optimal use of resources coincide, apart
from an initial transient, with those of maximal expansion
\cite{MacK}. Note that the $s_i$'s are essentially the discrete-time
version of the continuous time fluxes $\nu_i$ of Eq. (\ref{fba}).

It is intuitive that the solution space of (\ref{vn}) shrinks as
$\rho$ is increased starting from $\rho=0$, where any flux
configuration is a viable state. Quite importantly, then, for a fixed
set of input-output relations, the solutions of (\ref{vn}) are bound
to coincide with those of (\ref{fba}) if $\rho^\star=1$ and
$c^\mu=0\!\!\quad\forall\mu$.

The typical behaviour of Von Neumann's model can be fully appreciated
analytically in the case of random graphs.  Depending on their size
and topology, autocatalytic networks with random stoichiometry give
rise to very different optimal states. As in other
constrained optimization problems \cite{SAT}, the key control
parameter is the ratio $N/P=n$ of variables-to-constraints.  In random
topologies \cite{VNfc,VNd} as $n$ increases the system crosses over
from a contracting phase with $\rho^\star<1$ to an expanding one with
$\rho^\star>1$, passing through a stationary regime with
$\rho^\star=1$ in which reaction fluxes are constant in time as in FBA
(though the values at which fluxes settle may be different).

\section{Application to E. coli}

A natural reaction network can be modeled as an autocatalytic system
when the uptake reactions that provide resources and account for
exchanges of metabolites with the surrounding environment are
included. We have applied the Von Neumann scheme to the cellular
metabolism of the bacterium {\it E. coli}, as reconstructed in
\cite{Pals1}. To set the stage, three different operations have to be
performed.

(a) Environment selection. To fix the environment where the cell lives, we 
have defined a set of external metabolites on which we applied uptake 
fluxes. We have considered three distinct environmental conditions: 
(i) isolated cell, without uptakes; (ii) minimal environment, with uptakes 
on a restricted set of metabolites \cite{Bara1}, namely
CO$_2$, K, NH$_4$, PI, O$_2$, SO$_4$ and either one
of glu-L, succ or glc); (iii) rich environment, with uptakes on all
external metabolites. 

(b) ``Leaf-removal''. Once the network is built, one has to remove 
from the internal metabolites those that are never
produced, since the corresponding constraints are only satisfied by
taking $\rho=0$ or by applying a null flux to the corresponding
reactions (as is actually done in FBA). 

(c) Accounting for reversibility. We have disposed of reversible
reactions by introducing two separate fluxes and taking the absolute
difference as the positive net flux.

The size of the resulting network, i.e. $N$ and $P$, is ultimately
different for different environments.

We calculate the maximum growth rate $\rho^\star$ numerically via a
generalized MinOver algorithm \cite{Min}, detailed in \cite{VNd}, in
which solutions are found at fixed $\rho$ and then $\rho$ is gradually
increased. It has been shown rigorously \cite{VNd} that this algorithm
converges to a solution at a fixed $\rho$ if at least one solution
exists. Moreover, when multiple solutions occur (in which case they
form a connected set), the algorithm provides a uniform sampling of
the solution space \cite{sup}. Anticipating that indeed many flux
vectors satisfy (\ref{vn}) at $\rho^\star$ for {\it E. coli}, the
latter is a particularly important advantage since a uniform sampling
is required to characterize the solution space.

\section{Characterization of the solutions}

\begin{figure}
\begin{center}
\includegraphics*[width=10cm]{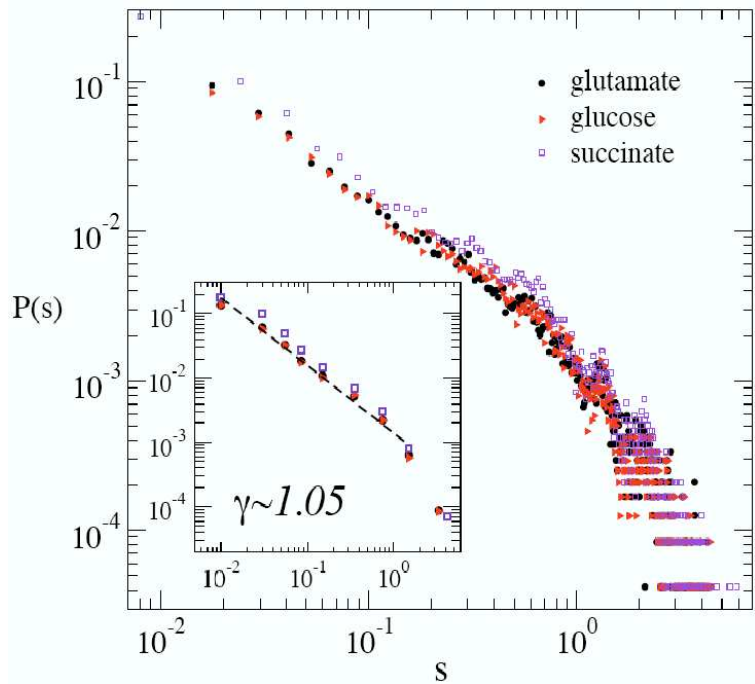}
\caption{\label{distr}Flux distributions at $\rho^\star$ for {\it
E. coli} in a minimal environment with three different carbon sources:
glutamate (glu), succinate (suc) and glucose (glc). Here, $N=1053$ and
$P=630$ (glu); $N=1054$ and $P=630$ (suc); $N=1057$ and $P=631$
(glc). Inset: the same distributions plotted after a binning of the
abscissa.}
\end{center}
\end{figure}

For the metabolic network of {\it E. coli}, we have found
$\rho^\star=0.999\pm 0.001$ independently of the environmental
conditions we have tested, so that the state of optimal growth is
compatible with one with constant fluxes. The distribution of reaction
fluxes at $\rho^\star$ (see Fig. \ref{distr}) displays a regime (less
than two decades) with a scaling form $P(s)\sim s^{-\gamma}$ with
exponent close to $-1$ (in reasonable agreement with the experimental
evidence based on a limited set of measured fluxes \cite{Emm,Bara1})
followed by a cutoff. Note that the flux histogram is remarkably
stable over different solutions. This scenario has been partially
reproduced by FBA \cite{Bara1}, but the solution obtained optimizing
the biomass production systematically overestimates $\gamma$. 

\begin{figure}
\begin{center}
\includegraphics*[width=10cm]{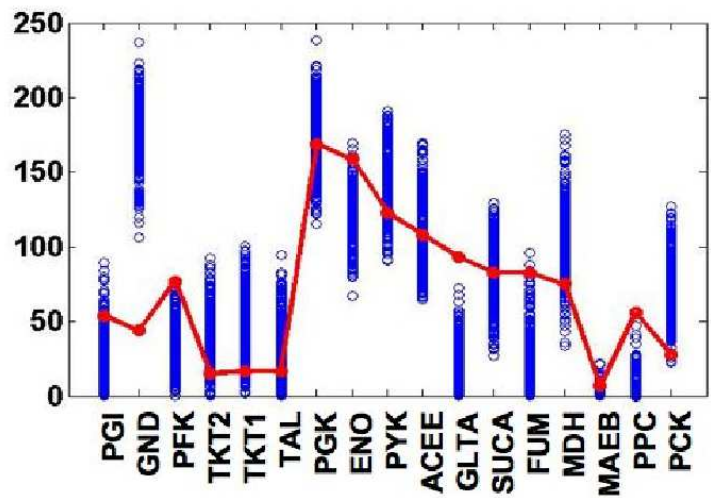}
\caption{\label{exps}Comparison of reaction fluxes predicted by the
Von Neumann scheme with 17 fluxes measured in \cite{Emm} and analyzed
in \cite{Segre}. The different reactions (in no specific order) are
reported on the horizontal axis (labeled as in \cite{Segre}), their
corresponding fluxes (relative to the glucose uptake) on the vertical
axis. Red markers denote experimental values, blue ones theoretical
predictions (see however text for details on the environmental
conditions considered). Note that many flux vectors satisfy Von
Neumann's conditions at optimal growth. 300 different solutions are
reported here.}
\end{center}
\end{figure}

In order to compare individual fluxes with experiments, we aimed at
studying the solution obtained in conditions similar to those
described in \cite{Emm}, focusing our attention on 17 fluxes from the
central metabolism, as in \cite{Segre}. Unfortunately, we are
unable to reproduce in detail the M9 medium used in \cite{Emm}, and
can only fix the uptakes of three metabolites, namely glucose, CO$_2$
and O$_2$ identically to \cite{Emm}. For the remaining part of the
environment, we chose to simulate a minimal medium and fixed the four
remaining external uptakes at arbitrary values (we did not observe a
strong dependency of the results on these parameters). Results are shown
in Fig. \ref{exps}. We stress that the medium we consider is not
strictly identical to that used in experiments. It is worth noting
(see also \cite{Sau2}) that experiments employ $^{13}$C-labeled carbon
sources as substrates for a growing bacterial culture kept at constant
density, and that the relative abundance of metabolites can be
captured by nuclear-magnetic-resonance or mass spectroscopy. Reaction
fluxes of different metabolic pathways are then inferred assuming a
model of the reaction network.

With stationarity recovered at $\rho^\star$=1, as discussed above the
difference between the Von Neumann solution and that of FBA arises
from the fact that not all $c^\mu$'s attain their lowest allowed
value. This is clearly seen in Fig. \ref{constr2}. In each solution,
some metabolites exist with a non-zero $c^\mu$, implying that in the
steady state a net production of such metabolites occurs. This may
also signal an incompleteness of the stoichiometric data.

\begin{figure}
\begin{center}
\includegraphics[width=10cm]{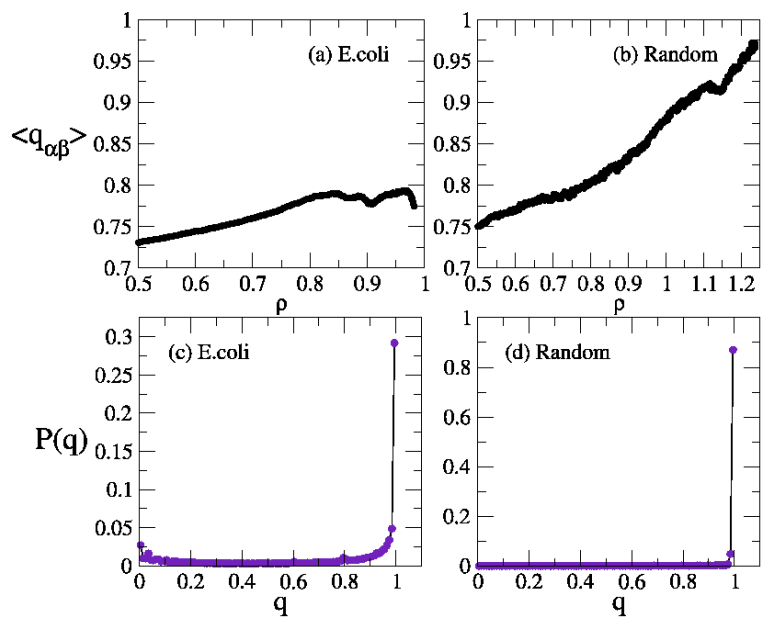}
\caption{\label{ov}(a,b) Mean overlap between 500 different solutions
as function of $\rho$ in {\it E. coli} and in a random metabolic
network; the last point on the abscissa corresponds to the value of
$\rho^\star$ estimated by simulation. (c,d) Overlap histogram $P(q)$
at $\rho^\star$ in E.coli and in the random network. Note the
different scales of the $y$ axes in the lower panels.}
\end{center}
\end{figure}

To get further insight on the existence of multiple flux states
compatible with (\ref{vn}) at $\rho^\star$ and on the shape of the
solution space, a rough but effective way consists in monitoring the
mean overlap between different solutions. Given two solutions $\alpha$
and $\beta$ at fixed $\rho$, we define their overlap $q_{\alpha
\beta}$ as:
\begin{equation}
q_{\alpha \beta}=\frac{2}{N}\sum_{i=1}^N\frac{
s_{i\alpha}s_{i\beta}}{s_{i\alpha}^2+s_{i\beta}^2}
=\frac{1}{N}\sum_{i=1}^N q_{\alpha\beta}^{(i)}\;.
\label{def.ov}
\end{equation}
This quantity equals one when ${\bf s}_\alpha$ and ${\bf s}_\beta$
coincide and becomes smaller and smaller as the distance between ${\bf
s}_\alpha$ and ${\bf s}_\beta$ in the space of flux vectors
increases\footnote{In computing it, one should consider a flux to be
zero whenever its value is below a threshold $\epsilon$, to take into
account the fact that there is a loss of information about relative
fluctuations between different solutions in fluxes smaller than
$\epsilon$. We have chosen $\epsilon=10^{-5}$ to ensure that overlaps
are not overestimated. Results obtained with $\epsilon=10^{-5},
10^{-6}$ and $10^{-7}$ are identical. We have furthermore taken into
account the fact that the overlap between null fluxes must be defined
by consistency to be equal to one.}. In Fig. \ref{ov} we report the
behaviour of the mean overlap $\langle q_{\alpha\beta}\rangle$ (the
angular brackets representing an average over many pairs of solutions)
for {\it E. coli} and for a network with the same topology and $N/P$
but random Gaussian stoichiometry. Random networks provide an
important benchmark against which the behaviour of real networks can
be tested, in order to highlight the extent to which observations are
specific of the particular organism that is being analyzed. In
particular, the role of topology and stoichiometry can be fully
analyzed. For instance, in \cite{VNd} it is shown that the power-law
behaviour of the flux distribution cannot be ascribed to the specific
network topology.

\begin{figure}
\begin{center}
\includegraphics[width=10cm]{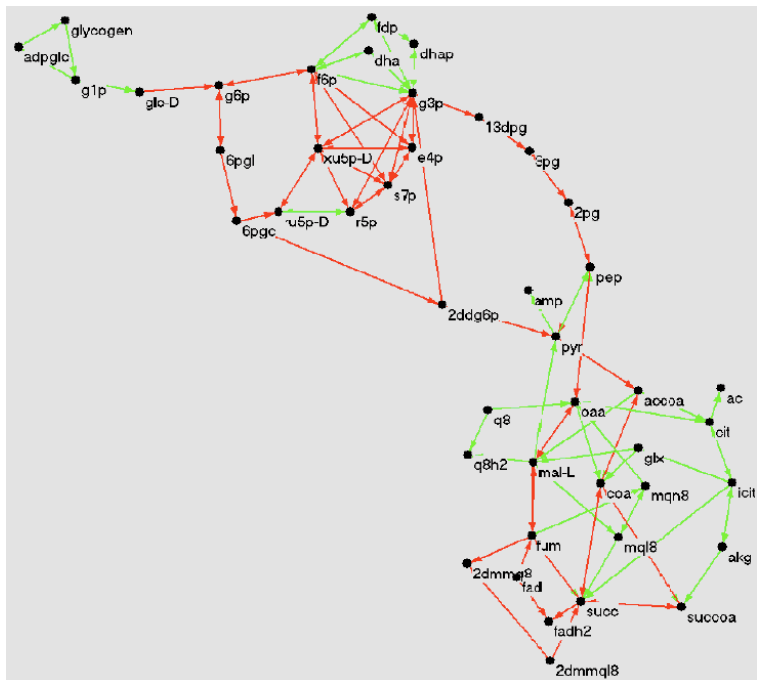}
\caption{\label{map}{\it E. coli}'s central metabolism: nodes
represent metabolites, an arrow joining two nodes is present when a
reaction exists converting one into the other. Red (resp. green) links
denote frozen (resp. free) reactions, with overlap larger
(resp. smaller) than $0.9$.}
\end{center}
\end{figure}

One sees that for the artificial metabolic network $\langle
q_{\alpha\beta}\rangle$ is an increasing function of $\rho$ that
approaches 1 as $\rho \to \rho^\star$ (Fig. \ref{ov}b). Moreover the
overlap histogram has a marked $\delta$-peak at $q=1$ whose mass
increases as $\rho\to\rho^\star$ (data not shown). These results
indicate that the optimal solution is unique.  On the other hand, the
behaviour of $\langle q_{\alpha\beta}\rangle$ for {\it E. coli}
suggests that the volume of solutions stops contracting when $\rho$
reaches roughly $0.8$ from below. In particular, at $\rho^\star$
multiple solutions survive.  From the corresponding histogram, Fig.
\ref{ov}c, we see that only about the $30\%$ of reactions have an
overlap close to 1. In a different jargon, one may say that roughly
$30\%$ of the variables are {\it frozen} (i.e. assume the same value
on all solutions of the constrained optimization problem), while the
remaining are {\it free}. The existence of frozen variables
characterizes many random constraint satisfaction problems \cite{gui},
but it is normally hard to identify topological motifs where variables
are more likely to be frozen. In the present case, it is reasonable to
expect that for purely structural reasons reaction chains are entirely
frozen when the first reaction of the chain is.  This is indeed
confirmed by the map of frozen/free fluxes in {\it E. coli}'s central
metabolism, displayed in Fig. \ref{map}. The chain-like part of
glycolisis appears indeed to be frozen.

\begin{figure} 
\begin{center}
\includegraphics[width=12cm]{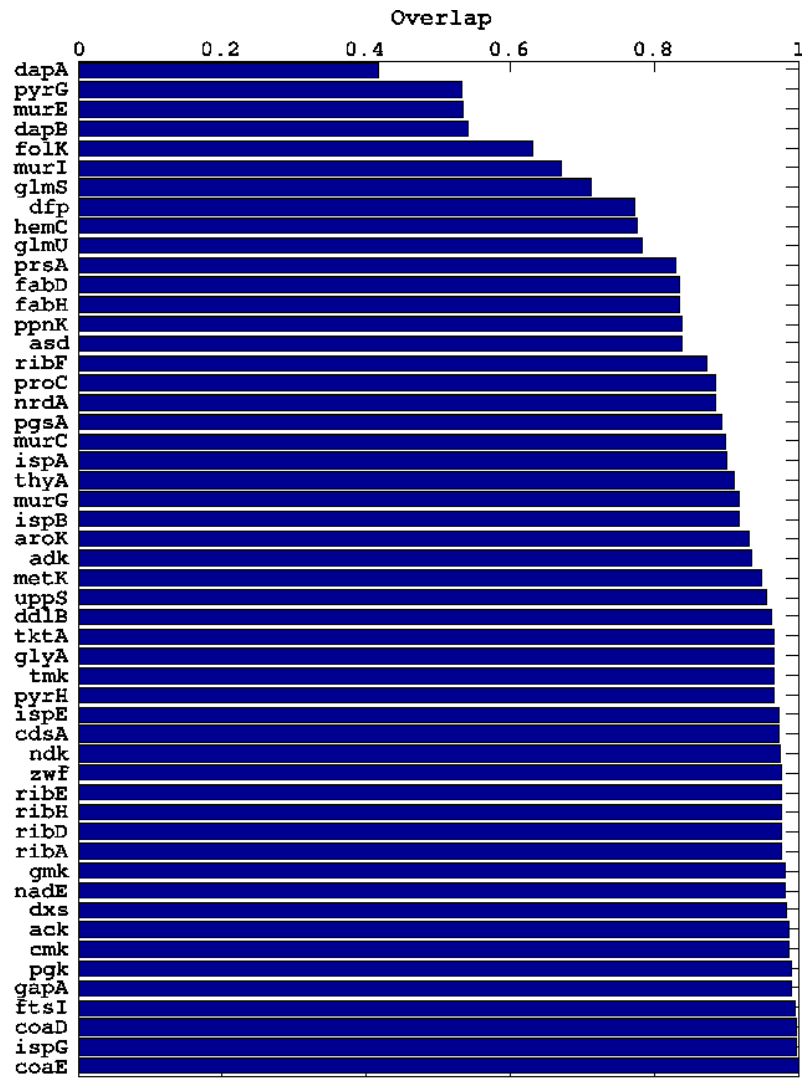}
\caption{\label{ci}Essential genes (vertical axis) versus overlap of
the corresponding reactions in the reconstructed metabolic network of
{\it E. coli} (horizontal axis).}
\end{center}
\end{figure}

The existence of frozen fluxes raises the obvious question of their
biological significance. To address this point, we have correlated
reaction overlaps $q_{\alpha\beta}^{(i)}$ with the essentiality of the
corresponding genes according to the notion of universal essentiality
used in \cite{gerdes}, which combines phenomenological relevance (a
gene is essential if knocking it out the cell dies) with evolutionary
retention (the presence of the gene in different species). In
\cite{gerdes}, 55 essential genes of {\it E. coli} involved in
metabolism have been identified, that are also present in 80\% of 32
different bacterial genomes. We have been able to link 52 of such
genes to reactions in the reconstructed network. It turns out, see
Fig. \ref{ci}, that 43 of such genes correspond to reactions with
overlap larger than $0.8$ and that only $7$ genes relate to reactions
with an overlap significantly smaller than 80\%\footnote{Whether a
gene is `essential' or not (according to the definition of
\cite{gerdes}) depends on the choice of the environment. In
particular, \cite{gerdes} considers a rich medium as the
environment. For comparison, we have considered theoretical values of
the overlaps for fluxes calculated assuming a rich environment. Note
that, in principle, the same gene can be linked to a more or less
variable flux in different environments. However, the results
presented are qualitatively preserved in the minimal environment with
different carbon sources.}. This suggests that frozen fluxes, which in
Von Neumann's framework are allowed a very limited variability if a
state of optimal growth is to be kept, may carry an evolutionary
significance.

\section{The role of conserved moieties}

\begin{figure}
\begin{center}
\includegraphics[width=10cm]{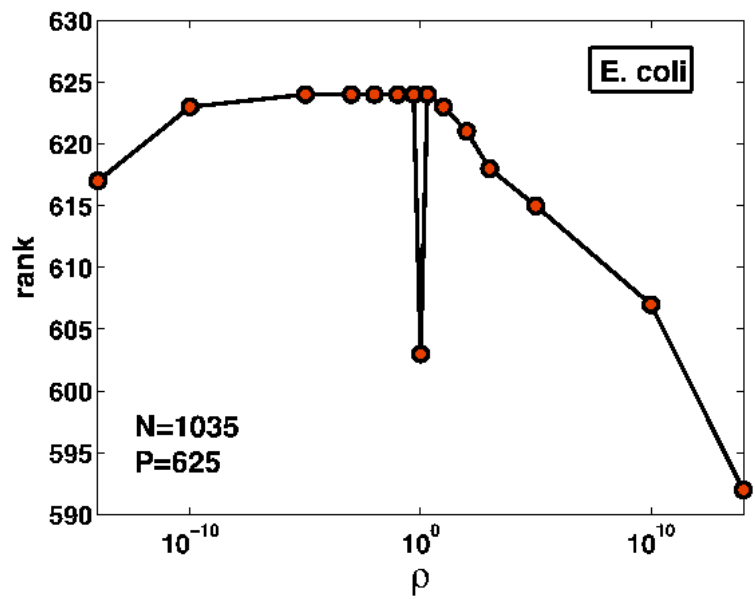}
\caption{\label{rnk}The rank of the matrix associated to the system of
inequalities (\ref{vn}) for {\it  E.Coli} (isolated cell) as a
function of $\rho$. Around the singularity the rank equals the
number of metabolites; conserved pools of metabolites are only present
exactly at $\rho=1$.}
\end{center}
\end{figure}

To trace back the physical origin of the results presented above, we
have studied the rank of the matrix ${\bf A}-\rho{\bf B}$ associated
to the $P$ linear constraints (\ref{vn}) as a function of the
parameter $\rho$, see Fig. \ref{rnk}.
The singularity occurring at $\rho=1$ is related to the presence of
conserved pools of metabolites, groups of reagents whose total
concentration is constant in time \cite{Pools}. Their existence is due
to the fact that the concentrations of metabolites of a certain pool
are always coupled with a common functional group. For example the
three metabolites ATP, AMP and ADP belong to a pool of cofactors that
preserve the adenylate moiety \cite{niko}. This implies that a change
in the concentration of a given metabolite can not be accomplished
without considering the entire pool to which that metabolite belongs.

A conserved pool $g$ is formally defined by a $P$-dimensional 
Boolean vector of elements $z^\mu_g$ such that $z^\mu_g=1$ 
if $\mu \in g$ and $z^\mu_g=0$ if $\mu \not\in g$ satisfying
\begin{equation}\label{g}
\sum_{\mu=1}^P z^\mu_g (a_i^\mu-b_i^\mu) = 0 \quad,\quad \forall
i=1,\ldots,N.
\end{equation}
Such conservation laws manifest themselves precisely at $\rho=1$, 
By virtue of (\ref{vn}), the relation $\sum_\mu z^\mu_g c^\mu \ge 0$
must hold for any $g$. It follows that either $\rho^\star\le 1$, or
all fluxes connected to a metabolite belonging to a conserved pool
must be equal to zero. Since the null solution $\mathbf{s}=\mathbf{0}$
must be discarded on obvious physical grounds, if all fluxes are
connected to metabolites in a conserved pool then necessarily
$\rho^\star\leq 1$.  This is consistent with the results obtained in
different environmental conditions and implies that this scenario must
be stable against small perturbations of the network topology.

\begin{figure} 
\begin{center}
\includegraphics[width=14cm]{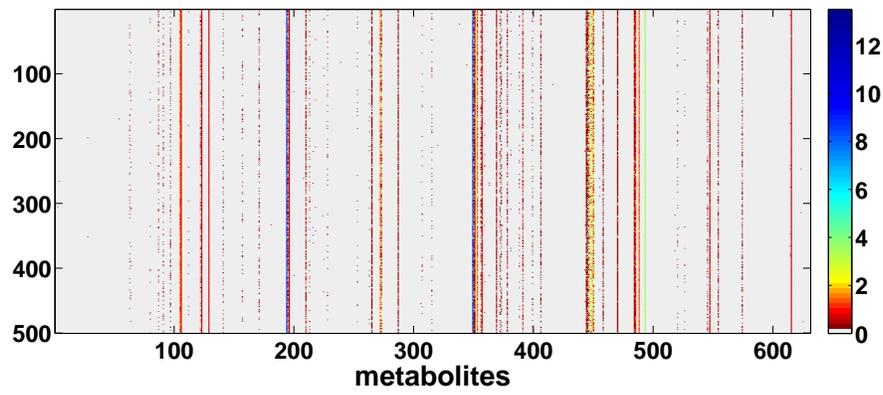}
\caption{\label{constr2}Values of $c^\mu$ for different metabolites
(horizontal axis) in 500 different solutions. The colour code is
reported on the right. The white background corresponds to
$c^\mu=0$. Colored marks indicate a net production of the
corresponding metabolite in that solution, or a mass balance
violation. One sees that while mass balance holds for most
metabolites, some are consistently unbalanced while others may or may
not be, depending on the solution.}
\end{center}
\end{figure}

Moreover, from (\ref{g}) and (\ref{vn}), it is easy to see that, for
$\rho=1$, $\mu\in g$ implies $c^\mu=0$, i.e. for a metabolite
belonging to a conserved pool the mass-balance condition must be
strictly valid. We have shown instead (Fig. \ref{constr2}) that the
values of the constraints are not always zero: for some metabolites,
the mass balance condition is not reached and we can assert that they
do not belong to any conserved pool (within the stoichiometric
description employed).
We conclude that in the state of optimal growth with
$\rho^\star=1$, in addition to the stationarity of reaction rates, a
spontaneous condition of mass balance holds for most, not all,
metabolites.

In summary, the Von Neumann model relies on the assumption that the
arrangement of metabolic fluxes follows a principle of growth
maximization. It does not imply {\it a priori} either the mass balance
or the stationarity of fluxes, but these two conditions are
essentially recovered at the maximum growth rate. This is due to the
presence of conserved metabolic pools. The singularity in the rank of
$\bf{A}-\rho\bf{B}$ at $\rho=1$ implies that the physically relevant
solutions are those to which the system tends as $\rho\to 1$. In this
limit, mass balance is recovered (i.e. $c^\mu=0$) for most but not all
metabolites. This provides a non-trivial correction to the picture
extracted by FBA and reproduces the limited experimental evidence that
is available.

\begin{acknowledgments}
We gratefully acknowledge important discussions 
with  G. Bianconi, R. Monasson, A. Pagnani and F. Ricci Tersenghi,
and a number of useful suggestions.
M.M. was supported by IST STREP GENNETEC contract no. 034952.
\end{acknowledgments}

\end{article}

\newpage

\newpage

\newpage

\newpage

\newpage

\end{document}